\documentclass[12pt]{article}
\usepackage{epsf}
\title{ {\bf The possible effects of non-universal extra dimensions on
$t\rightarrow c \, l_i^- l_j^+ $ and $H^0 \rightarrow h^0 (A^0)
l_i^- l_j^+$ decays in the general two Higgs Doublet model}}
\author{\vspace{1cm}\\
        {\bf E. O. Iltan}
        \thanks{E-mail address:
        eiltan@heraklit.physics.metu.edu.tr}
 \\
        Physics Department, Middle East Technical University \\
        Ankara, Turkey\\}

\date{}

\begin{document}
\setlength{\baselineskip}{24pt}
\maketitle
\setlength{\baselineskip}{7mm}
\begin{abstract}
We study the effects of non-universal extra dimensions on the
decay widths of the lepton flavor violating  processes,
$t\rightarrow c \, l_i^- l_j^+ $ and $H^0 \rightarrow h^0 (A^0)
l_i^- l_j^+$ in the general two Higgs doublet model. We consider
that the extra dimensions are accessible to the standard model
gauge fields and the new Higgs doublet. We observe that  the
lepton flavor violating $H^0 \rightarrow h^0 (\tau^+ \mu^- +
\tau^- \mu^+)$ and $H^0 \rightarrow A^0 (\tau^+ \mu^- + \tau^-
\mu^+)$ Higgs decays are sensitive to the extra dimensions,
especially, in the case of two spatial ones. This result may
ensure a test to determine the compactification scale and the
possible number of extra dimensions, with the accurate
experimental measurements.
\end{abstract}
\thispagestyle{empty}
\newpage
\setcounter{page}{1}
\section{Introduction}
Lepton flavor violating (LFV) interactions are worthwhile to study
since they ensure comprehensive information about the possible new
physics effects beyond the standard model (SM) and the free
parameters existing in these models. Such processes occur with the
help of the tree level flavor changing neutral currents (FCNC),
which appear in the models beyond the SM, like the two Higgs
doublet model (2HDM). It is well known that the model III version
of the 2HDM possesses the FCNC at tree level and the strengths of
the flavor changing (FC) interactions are regulated by the Yukawa
couplings, appearing as free parameters which should be restricted
by the experimental data. The addition of extra dimensions into
the theory brings new extension and the search of the effects of
the possible new dimensions would be illustrative to test their
existence.

Extra dimensions are introduced for solving the gauge hierarchy
problem of the SM and there are various studies on this subject in
the literature \cite{Appelquist}-\cite{Lam}. The idea is that
there is a fundamental theory lying in higher dimensions and the
ordinary four dimensional SM is its low energy effective theory.
This is achieved by considering that the extra dimension (two
extra dimensions) over four dimensions is compactified on orbifold
$S^1/Z_2$ ($(S^1\times S^1)/Z_2$) with small radius $R$, which is
a typical size of the extra dimension(s). This compactification
results in the production of Kaluza-Klein (KK) states of the
fields with masses regulated by the parameter $R$. If the extra
dimensions are accessible to all fields in the model, they are
called as universal extra dimensions (UED) in the literature
\cite{Appelquist}-\cite{Appelquist2}. In this case, the extra
dimensional momentum is conserved at each vertex and the
interactions with only one KK state are forbidden, i.e., the KK
number is conserved. The conservation of the KK number leads to
the appearance of heavy stable particles. Furthermore, this
conservation causes that the KK modes enter into the calculations
as loop corrections and, therefore, the constraints on the size of
the extra dimensions which are obtained from SM precision
measurements are less stringent. The size of compactification
scale has been studied by taking into account the loop effects
induced by the internal top quark and it has been estimated in the
range of $200-500\, GeV$, using the electroweak precision
measurements \cite{Appelquist}, the \( B-\bar{B} \) -mixing
\cite{Papavassiliou},\cite{Chakraverty} and the flavor changing
process $b \to s \gamma$ \cite{Agashe}. In several works
\cite{Arkani, Antoniadis1,Antoniadis2,Carone,Antoniadis3}, this
size has been estimated as large as few hundreds of GeV. In the
case of non-universal extra dimensions, where some of the
particles are confined on 4D brane and do not feel the new
dimensions, the coupling of two zero modes with the KK mode is
permitted and this ensures to predict the effects of extra
dimensions even at tree level.

Our work is devoted to the analysis of the LFV t-quark
$t\rightarrow c \, l_i^- l_j^+ $ and  Higgs boson $H^0 \rightarrow
h^0 l_i^- l_j^+$ and $H^0 \rightarrow A^0 l_i^- l_j^+$ decays in
the framework of the model III, with the addition of a single
(two) extra dimension(s). We consider that the new Higgs doublet
and the gauge sector of the SM feel the extra dimensions, however,
the other SM fields are confined on 4D brane. Since these decays
can exist at tree level in the model III, the higher dimensional
effects for non-universal case under consideration appear with the
intermediate (virtual) neutral Higgs fields $H$, namely $h^0$ and
$A^0$, which can create "two zero modes-KK mode" vertices, in
contrast to the case of UED. In the present analysis, we try to
predict the additional effects due to a single and two spatial
extra dimensions. In the case of a single extra dimension, the KK
modes of the neutral Higgs fields $H$, with masses
$\sqrt{m_{H}^2+n^2/R^2}$, appear after the compactification on
orbifold $S^1/Z_2$. Here $m_n=n/R$ is the mass of $n$'th level KK
particle where $R$ is the compactification radius. If there exist
two spatial extra dimensions which are accessible to the new Higgs
doublet, the non-zero KK modes of the neutral Higgs fields $H$
have the masses $\sqrt{m_{H}^2+m_n^2+m_k^2}$, where the mass terms
$m_n=n/R$ and $m_k=k/R$ are due to the compactification of the
extra dimensions on orbifold $(S^1\times S^1)/Z_2$
\cite{Appelquist}.

In the numerical calculations we see that the extra dimension
contribution to the  FV $t\rightarrow c\, (\tau^+ \mu^- + \tau^-
\mu^+)$ decay is negligible, at least, up to two extra dimensions.
However, the LFV $H^0 \rightarrow h^0 \, (\tau^+ \mu^- + \tau^-
\mu^+)$ and $H^0 \rightarrow A^0 \, (\tau^+ \mu^- + \tau^- \mu^+)$
Higgs decays are sensitive to the extra dimensions and the
predictions of additional effects to their decay widths are almost
comparable with the decay widths obtained without extra
dimensions, in the case of two extra dimensions. This result may
ensure a test to determine the compactification scale and the
possible number of extra dimensions.

The paper is organized as follows: In Section 2, we present the
decay widths of LFV interactions $t\rightarrow c \, (l_i^- l_j^+
+l_j^- l_i^+) $ and  $H^0 \rightarrow h^0 (A^0) (l_i^- l_j^+
+l_j^- l_i^+)$  in the model III version of the 2HDM, with the
inclusion of non-universal extra dimensions. Section 3 is devoted
to discussion and our conclusions.
\section{The LFV interactions $t\rightarrow c \, (l_i^- l_j^+
+l_j^- l_i^+) $ and  $H^0 \rightarrow h^0 (A^0) (l_i^- l_j^+
+l_j^- l_i^+)$  in the general two Higgs Doublet model with the
inclusion of non-universal extra dimensions}
The FV interactions are worthwhile to investigate and, among them,
the LFV interactions receive great interest since the theoretical
predictions of their branching ratios (BR's) in the framework of
the SM are small and this forces one to go the new models beyond.
The model III version of the 2HDM, permitting tree level neutral
currents, is one of the candidate that can ensure additional
contributions to the physical quantities with the appropriate
choice of free parameters, such as Yukawa couplings, masses of new
particles. The inclusion of spatial extra dimensions causes to
enhance the BR's of these decays and these enhancements depend on
the compactification scale $1/R$, where $R$ is the radius of the
compactification.

Now, we assume that the second Higgs doublet feels the extra
dimensions. We start with the part of the Lagrangian which is
responsible for the FV vertex, the so called Yukawa Lagrangian, in
a single extra dimension:
\begin{eqnarray}
{\cal{L}}_{Y}&=&\eta^{U}_{ij} \bar{Q}_{i L} \tilde{\phi_{1}} U_{j
R}+ \eta^{D}_{ij} \bar{Q}_{i L} \phi_{1} D_{j R}+
\xi^{U\,\dagger}_{5 \,ij} \bar{Q}_{i L} (\tilde{\phi_{2}}|_{y=0})
U_{j R}+ \xi^{D}_{5 \,ij} \bar{Q}_{i L} (\phi_{2}|_{y=0}) D_{j R}
\nonumber \\
&+& \eta^{E}_{ij} \bar{l}_{i L} \phi_{1} E_{j R}+ \xi^{E}_{5 \,ij}
\bar{l}_{i L} (\phi_{2}|_{y=0}) E_{j R} + h.c. \,\,\, ,
\label{lagrangian}
\end{eqnarray}
where $y$ represents the extra dimension, $L$ and $R$ denote
chiral projections $L(R)=1/2(1\mp \gamma_5)$, $\phi_{i}$ for
$i=1,2$, are two scalar doublets, $\bar{Q}_{i L}$ are left handed
quark doublets, $U_{j R} (D_{j R})$ are  right handed up (down)
quark singlets, $l_{i L}$ ($E_{j R}$) are lepton doublets
(singlets), with family indices $i,j$. The Yukawa couplings
$\xi^{E}_{5\, ij}$ are dimensionful and rescaled to the ones,
$\xi^{U,D,E}_{ij}$, in four dimensions as $\xi^{U,D,E}_{5\,
ij}=\sqrt{2 \pi R}\,\, \xi^{U,D,E}_{ij}$.

In the present work, we assume that the Higgs doublet lying in the
four dimensional brane has non-zero vacuum expectation value which
ensures the ordinary masses of the gauge fields and the fermions.
On the other hand, the second doublet, which is accessible to the
extra dimension, does not receive the vacuum expectation value.
Namely, we choose the doublets $\phi_{1}$ and $\phi_{2}$ and the
their vacuum expectation values as
\begin{eqnarray}
\phi_{1}=\frac{1}{\sqrt{2}}\left[\left(\begin{array}{c c}
0\\v+H^{0}\end{array}\right)\; + \left(\begin{array}{c c} \sqrt{2}
\chi^{+}\\ i \chi^{0}\end{array}\right) \right]\, ;
\phi_{2}=\frac{1}{\sqrt{2}}\left(\begin{array}{c c} \sqrt{2}
H^{+}\\ H_1+i H_2 \end{array}\right) \,\, , \label{choice}
\end{eqnarray}
and
\begin{eqnarray}
<\phi_{1}>=\frac{1}{\sqrt{2}}\left(\begin{array}{c c}
0\\v\end{array}\right) \,  \, ; <\phi_{2}>=0 \,\, .\label{choice2}
\end{eqnarray}
With the choice under consideration the mixing between neutral
scalar Higgs bosons is switched off and it would be possible to
separate the particle spectrum so that the SM particles are
collected in the first doublet and the new particles in the second
one. Here $H_1$ ($H_2$) is the well known mass eigenstate $h^0$
($A^0$). Notice that both Higgs doublets can have non-zero vacuum
expectation values in general and this leads to the mixing between
the neutral Higgs bosons $H^0$ and $h^0$ in the CP even sector. In
the CP odd one, the mixing appears between $\chi^{0}$ and $H_2$.
There exist new parameters which include the mixing angle of CP
even neutral Higgs bosons and the ratio of the vacuum expectation
values of each Higgs doublet, in the vertices (for example
$H^0-h^0-h^0$ and $H^0-A^0-A^0$ vertices, lepton-lepton Higgs
boson vertices (see \cite{Kane} and \cite{Atwood} for review).
Therefore, in general, the mixing angle and the ratio of the
vacuum expectation values appear in the physical quantities.
%
%

The part which produce FCNC at tree level
\begin{eqnarray}
{\cal{L}}_{Y,FC}&=& \xi^{U\,\dagger}_{5 \,ij} \bar{Q}_{i L}
(\tilde{\phi_{2}}|_{y=0}) U_{j
R}+ \xi^{D}_{5 \,ij} \bar{Q}_{i L} (\phi_{2}|_{y=0}) D_{j R} \nonumber \\
&+& \xi^{E}_{5 \,ij} \bar{l}_{i L} (\phi_{2}|_{y=0}) E_{j R} +
h.c. \,\, ,\label{lagrangianFC}
\end{eqnarray}
carries the information about the extra dimension over the second
Higgs doublet $\phi_{2}$ and it can be expanded into its KK modes
after the compactification of the extra dimension on orbifold
$S^1/Z_2$ as
\begin{eqnarray}
\phi_{2}(x,y ) & = & {1 \over {\sqrt{2 \pi R}}} \left\{
\phi_{2}^{(0)}(x) + \sqrt{2} \sum_{n=1}^{\infty} \phi_{2}^{(n)}(x)
\cos(ny/R)\right\} \, , \label{SecHiggsField}
\end{eqnarray}
where $\phi_{2}^{(0)}(x)$ is the four dimensional Higgs doublet
which contains the charged Higgs boson $H^+$, the neutral CP even
(odd) $h^0$ ($A^0$) Higgs bosons and $R$ is the compactification
radius.
Furthermore, each non-zero KK mode of Higgs doublet $\phi_{2}$
includes a charged Higgs of mass $\sqrt{m_{H^\pm}^2+m_n^2}$, a
neutral CP even Higgs of mass $\sqrt{m_{h^0}^2+m_n^2}$, a neutral
CP odd Higgs of mass $\sqrt{m_{A^0}^2+m_n^2}$ where $m_n=n/R$ is
the mass of $n$'th level KK particle, emerging from
compactification.

Now, we start to investigate the LFV inclusive  $t\rightarrow c \,
l_i^- l_j^+ $ decay  where $l_i,\, l_j$ are different lepton
flavors (see Fig. 1) in the model III, where only the new Higgs
doublet feels a single extra dimension. This process can exist at
tree level, by taking non-zero $t-c\, (l_i^-  l_j^+)$ transition
driven by the neutral bosons $h^0$ and $A^0$. There are FV vertex
in the quark sector, $t-c\, h^{0 *} (A^{0 *})$ and it is connected
to the $l_i^- l_j^+$ outgoing leptons. Since only the new Higgs
doublet, and therefore, the $h^0$ and $A^0$ bosons, feels extra
dimension, the KK modes of them contribute to the process in
addition to their zero modes (see Fig. 1). Notice that in the case
of UED there would be no contribution coming from the extra
dimension at tree level due to the KK number conservation.

Here we present the matrix element square of the process
$t\rightarrow c\, (l_i^- l_j^+ + l_j^+ l_i^-)$ (see
\cite{Iltantcl1l2})
\begin{eqnarray}
|M|^2&=& 8\,m_t^2\, (1-s) \, \sum_{H=h^0,A^0} |p_H|^2\, \Big(
|a^{(q)}_{H}|^2+|a^{\prime \, (q)}_{H}|^2 \Big)\, \Big( (s\, m_t^2
- (m_{l_i^-}-m_{l_j^+})^2 )\, |a^{(l)}_{H}|^2 \nonumber \\ &+&
(s\, m_t^2 - (m_{l_i^-}+m_{l_j^+})^2)\, |a^{\prime \, (l)}_{H}|^2
\Big) \,
\nonumber \\
&+& 16\, m_t^2\, (1-s) \,\Bigg( (s\, m_t^2 -
(m_{l_i^-}-m_{l_j^+})^2)\, Re [p_{h^0}\,p^*_{A^0}\,
a^{(l)}_{h^0}\, a^{*(l)}_{A^0}\, (a^{(q)}_{h^0}\, a^{*
(q)}_{A^0}+a^{\prime \, (q)}_{h^0}\, a^{\prime \,  * (q)}_{A^0})]
\nonumber \\ &+& (s\, m_t^2 - (m_{l_i^-}+m_{l_j^+})^2)\, Re
[p_{h^0}\,p^*_{A^0}\, a^{\prime \, (l)}_{h^0}\, a^{\prime \, *
(l)}_{A^0}\, (a^{(q)}_{h^0}\, a^{* (q)}_{A^0}+ a^{\prime \,
(q)}_{h^0}\, a^{\prime \, * (q)}_{A^0})] \Bigg ) \, ,
\label{M2tot}
\end{eqnarray}
where
\begin{eqnarray}
p_H=\frac{i}{k^2-m^2_H+i m_H\,\Gamma^{H}_{tot}} +2 \,
\sum_{n=1}^{\infty} \frac{i}{k^2-m^2_{H^n}}\, , \label{pH}
\end{eqnarray}
and $\Gamma^{H}_{tot}$ is the total decay width of $H$ boson, for
$H=h^0,\, A^0$. In eq. (\ref{pH}), the parameter $s$ is
$s=\frac{k^2}{m_t^2}$, with the intermediate $H$ boson momentum
square $k^2$ and $m_{H^n}$ is the mass of $n^{th}$ KK mode of $H$
boson, $m_{H^n}=\sqrt{m^2_{H}+\frac{n^2}{R^2}}$. Here the
functions $a^{(l)}_{h^0, A^0}$, $a^{\prime \,(l)}_{h^0, A^0}$
read,
\begin{eqnarray}
a^{(l)}_{h^0}&=&-\frac{i}{2\sqrt{2}}\, (\xi^D_{N,l_i l_2}+\xi^{*
D}_{N,l_2 l_i})
\, , \nonumber  \\
a^{ (l)}_{A^0}&=&\frac{1}{2\sqrt{2}}\, (\xi^D_{N,l_i l_j}-\xi^{*
D}_{N,l_j l_i})
\, , \nonumber  \\
a^{\prime\,  (l)}_{h^0}&=&-\frac{i}{2\sqrt{2}}\, (\xi^D_{N,l_i
l_j}-\xi^{* D}_{N,l_j l_i})
\, , \nonumber  \\
a^{\prime\,  (l)}_{A^0}&=&\frac{1}{2\sqrt{2}}\, (\xi^D_{N,l_i
l_j}+\xi^{* D}_{N,l_j l_i})
\, , \nonumber  \\
a^{ (q)}_{h^0}&=&\frac{i}{2\sqrt{2}}\, (\xi^U_{N,tc}+\xi^{*
U}_{N,ct})
\, , \nonumber  \\
a^{ (q)}_{A^0}&=&-\frac{1}{2\sqrt{2}}\, (\xi^U_{N,tc}-\xi^{*
U}_{N,ct})
\, , \nonumber  \\
a^{\prime\,  (q)}_{h^0}&=&\frac{i}{2\sqrt{2}}\,
(\xi^U_{N,tc}-\xi^{* U}_{N,ct})
\, , \nonumber  \\
a^{\prime\,  (q)}_{A^0}&=&-\frac{i}{2\sqrt{2}}\,
(\xi^U_{N,tc}+\xi^{* U}_{N,ct}) \, .  \label{aql3}
\end{eqnarray}
Notice that we replace $\xi^{U,D,E}$ with $\xi^{U,D,E}_{N}$ where
"N" denotes the word "neutral". Using the eq. (\ref{M2tot}), the
differential decay width (dDW) $\frac{d\Gamma}{ds}(t\rightarrow c
\, (l_1^- l_2^++l_1^+ l_2^-) )$ is obtained as
\begin{eqnarray}
\frac{d\Gamma}{ds}=\frac{1}{256\,N_c\,\pi^3}\,\lambda\, |M|^2 \, ,
\end{eqnarray}
where $\lambda$ is: \\
$\lambda=\frac{\sqrt{\Big(m_t^2\,(s-1)^2-4\,m_c^2 \Big) \, \Big(
m_c^4+m_{l_i}^4+(m_{l_j}^2-m_t^2\,s)^2-2\,m_c^2\,(m_{l_i}^2+
m_{l_j}^2-m_t^2\,s)-2\,m_{l_i}^2\,(m_{l_j}^2+m_t^2\,s)  \Big)}}
{2\,m_t^2\,s}$. Here the parameter $s$ is restricted into the
region $\frac{(m_{l_i}+m_{l_j})^2}{m_t^2}\leq s \leq
\frac{(m_t-m_c)^2}{m_t^2}$.

At this stage we study the processes $H^0 \rightarrow h^0 l_i^-
l_j^+$ and $H^0 \rightarrow A^0 l_i^- l_j^+$ (see \cite
{IltanH0h0A0l1l2}) where $l_i,\, l_j$ are different lepton flavors
(see Fig. \ref{fig2}) and we consider the model III version of the
2HDM with the addition of extra dimension that is felt by the new
Higgs doublet, similar to the previous calculation. These
processes exist at tree level and the extra dimension effects
appear in the case of virtual $h^0 \, (A^0)$ transitions (see Fig.
\ref{fig2}-c and \ref{fig2}-d). The KK modes of these neutral
Higgs bosons contribute to the processes contrary to the UED case
where there would be no contribution coming from the extra
dimension at tree level.

Using the diagrams Fig. \ref{fig2}-a and Fig. \ref{fig2}-c the
matrix element square of the process $H^0 \rightarrow h^0 l_i^-
l_j^+$ reads
\begin{eqnarray}
|M|^2=A_1+A_2+A_3 \,, \nonumber \\
\label{M2h0ij}
\end{eqnarray}
where
\begin{eqnarray}
 A_1&=&\frac{1}{2 (m^2_{H^0}+2 p.k_{l_i})^2}\, \, \Bigg\{
m^2_{l_i}\, |\xi^{E}_{N,j i}|^2 \, \Bigg( 2
(p.k_{l_i})^2+(m^2_{H^0}- 4
m^2_{l_i})\,q.k_{l_i}+p.k_{l_i}\,(m^2_{H^0}\nonumber \\ &+& 4
m_{l_i}(m_{l_j}+2\,m_{l_i}-2 m_{l_j}\, sin^2\theta_{ij}) - 2 p.q)+
m_{l_i}(4\, m^2_{l_i}(m_{l_i}+m_{l_j}-2\,m_{l_j}\,
sin^2\theta_{ij})\nonumber
\\ &+& m^2_{H^0} (3\,m_{l_i}+m_{l_j}-2\,m_{l_j}\,
sin^2\theta_{ij})-4\,m_{l_j}\, p.q \Bigg)\Bigg\} \, , \nonumber
\\
A_2&=&\frac{1}{\sqrt{2} (m^2_{H^0}+2 p.k_{l_i})}\, \, \Bigg\{ 4\,
m_{l_i}\,m^2_{h^0}\, |\xi^{E}_{N,j i}|^2 \,Im[p_{h^0}] \Bigg( (3
m_{l_i}+m_{l_j}-2\,m_{l_j}\,sin^2\theta_{ij})\,p.k_{l_i} \nonumber
\\ &+& m_{l_i}\,(m^2_{H^0}+ 2\, m^2_{l_i}+2 m_{l_i} m_{l_j}-4\,
m_{l_i} m_{l_j}\,sin^2\theta_{ij} -2 \,q.(k_{l_i}-p))\Bigg)
\Bigg\}\, , \nonumber
\\
A_3&=& 4 m^4_{h^0}\, |\xi^{E}_{N,j i}|^2 \,Abs[p_{h^0}]^2 \Bigg(
m_{l_i}(m_{l_i}+m_{l_j}-2\,m_{l_j}\,sin^2\theta_{ij})+(p-q).k_{l_i}
\Bigg) \, , \label{M2h0ijA}
\end{eqnarray}
and
\begin{eqnarray}
p_{h^0}=\frac{i}{k^2-m^2_{h^0}+i m_{h^0}\,\Gamma^{{h^0}}_{tot}}
+2\, \sum_{n=1}^{\infty} \frac{i}{k^2-m^2_{{h^0}^n}}\, ,
\label{pS2}
\end{eqnarray}
with the transfer momentum square $k^2$, four momentum of incoming
$H^0$, outgoing $h^0$, outgoing $l_i^-$ lepton, $p$, $q$,
$k_{l_i}$, respectively. In eq. (\ref{M2h0ijA}), the parameter
$\theta_{ij}$ carries the information about the complexity of the
Yukawa coupling $\xi^{E}_{N,ij}$ with the parametrization
\begin{equation}
\xi^{E}_{N,i j}=|\xi^{E}_{N,ij}|\, e^{i \theta_{ij}} \, .
\label{xicomplex}
\end{equation}
Similarly, using the diagrams Fig. \ref{fig2}-b and Fig.
\ref{fig2}-d, the matrix element square of the process $H^0
\rightarrow A^0 l_i^- l_j^+$ is obtained as
\begin{eqnarray}
|M|^2=A'_1+A'_2+A'_3 \, , \nonumber \\
\label{M2A0ij}
\end{eqnarray}
where
\begin{eqnarray}
A'_1&=& \frac{1}{2 (m^2_{H^0}+2 p.k_{l_i})^2}\, \, \Bigg\{
m^2_{l_i}\, |\xi^{E}_{N,j i}|^2 \, \Bigg( 2
(p.k_{l_i})^2+(m^2_{H^0}-4
m^2_{l_i})\,q.k_{l_i}+p.k_{l_i}\,(m^2_{H^0}\nonumber \\ &+&
4\,m_{l_i}(-m_{l_j}+2\,m_{l_i}+2 m_{l_j}\, sin^2\theta_{l_i l_j})-
2 p.q)+ m_{l_i}(4\, m^2_{l_i}(m_{l_i}-m_{l_j}+2\,m_{l_j}\,
sin^2\theta_{ij})\nonumber\\ &+&
m^2_{H^0}(3\,m_{l_i}-m_{l_j}+2\,m_{l_j}\, sin^2\theta_{l_i
l_j})-4\,m_{l_j}\, p.q \Bigg)\Bigg\} \, ,
\nonumber \\
A'_2&=&\frac{1}{\sqrt{2} (m^2_{H^0}+2 p.k_{l_i})}\, \, \Bigg\{ 4\,
m_{l_i}\,m^2_{A^0} |\xi^{E}_{N,j i}|^2 \,Im[p_{A^0}] \Bigg( (3
m_{l_i}-m_{l_j}+2\,m_{l_j}\,sin^2\theta_{ij}) p.k_{l_i}\nonumber \\
&+& m_{l_i}\,(m^2_{H^0}+ 2\, m^2_{l_i}-2\, m_{l_i} m_{l_j}+4\,
m_{l_i} m_{l_j}\,sin^2\theta_{ij} -2 \,q.(k_{l_i}-p))\Bigg)
\Bigg\}\, , \nonumber
\\
A_3&=& 4\, m^4_{A^0}\, |\xi^{E}_{N,j i}|^2 \,Abs[p_{A^0}]^2 \Bigg(
m_{l_i}(m_{l_i}-m_{l_j}+2\,m_{l_j}\,sin^2\theta_{ij})+(p-q).k_{l_i}
\Bigg)\, , \label{M2A0ijA}
\end{eqnarray}
and $q$ is four momentum of outgoing $A^0$.
The decay width $\Gamma$ is obtained in the $H^0$ boson rest frame
by using the well known expression
\begin{equation}
d\Gamma=\frac{(2\, \pi)^4}{m_{H^0}} \, |M|^2\,\delta^4
(p-\sum_{i=1}^3 p_i)\,\prod_{i=1}^3\,\frac{d^3p_i}{(2 \pi)^3 2
E_i} \, , \label{DecWidth}
\end{equation}
where $p$ ($p_i$, i=1,2,3) is four momentum vector of $H^0$ boson,
($h^0$ ($A^0$) boson, outgoing $l_i^-$ and $l_j^+$ leptons).

Finally, we would like to analyze these decays in the two extra
spatial dimensions. With the assumption that the second Higgs
doublet $\phi_{2}$ feels the extra dimensions, it can be expanded
into its KK modes after the compactification of the extra
dimensions on orbifold $(S^1\times S^1)/Z_2$ as
\begin{eqnarray}
\phi_{2}(x,y,z ) & = & {1 \over {2 \pi R}} \left\{
\phi_{2}^{(0,0)}(x) + 2 \sum_{n,k} \phi_{2}^{(n,k)}(x) \cos(n
y/R+kz/R)\right\} \, , \label{SecHiggsField2}
\end{eqnarray}
where each circle is considered having the same radius $R$. In the
summation, the indices  $n$ and $k$ are positive integers
including zero but both are not zero at the same time. Here
$\phi_{2}^{(0,0)}(x)$ is the four dimensional Higgs doublet which
contains the charged Higgs boson $H^+$, the neutral CP even (odd)
$h^0$ ($A^0$) Higgs bosons. Each non-zero KK mode of Higgs doublet
$\phi_{2}$ includes a charged Higgs of mass
$\sqrt{m_{H^\pm}^2+m_n^2+m_k^2}$, a neutral CP even Higgs of mass
$\sqrt{m_{h^0}^2+m_n^2+m_k^2}$, a neutral CP odd Higgs of mass
$\sqrt{m_{A^0}^2+m_n^2+m_k^2}$ where the mass terms $m_n=n/R$ and
$m_k=k/R$ exist due to the compactification.

In the decays we consider that there appear KK modes $h^{0\,n,k}$
and $A^{0\,n,k}$ on the virtual $h^0$ and $A^0$ lines and the
parameter $p_H$ in eq. (\ref{M2tot}) (eqs. (\ref{M2h0ijA}) and
(\ref{M2A0ijA}) ) is redefined as
\begin{eqnarray}
p_H=\frac{i}{s\, m_t^2-m^2_H+i m_S\,\Gamma^{H}_{tot}} +
4\,\sum_{n,k} \frac{i}{s\, m_t^2-m^2_{H^{n,k}}}\, ,
\label{pS2Extr}
\end{eqnarray}
where $m_{H^{n,k}}=\sqrt{m_H^2+\frac{n^2+k^2}{R^2}}$.
\section{Discussion}
The LFV $t\rightarrow c \, l_i^- l_j^+ $ and $H^0 \rightarrow h^0
(A^0) l_i^- l_j^+$  decays exist at tree level in the model III
and the Yukawa couplings $\bar{\xi}^{U,D,E}_{N,ij}$ \footnote{We
use the parametrization $\xi^{U,D,E}_{N,ij}= \sqrt{\frac{4
G_F}{\sqrt {2}}} \bar{\xi}^{U,D,E}_{N,ij}$ for the Yukawa
couplings.}, with different quark and lepton flavors $i,j$, play
the main role in the interactions. Since these couplings are free
parameters of the theory, they need to be restricted by using the
experimental results. Now we will present the assumptions and the
numerical values we use for the free parameters under
consideration:
\begin{itemize}
\item The Yukawa couplings $\bar{\xi}^{U,D,E}_{N,ij}$ are
symmetric with respect to the indices $i$ and $j$.
\item The couplings $\bar{\xi}^{E}_{N,ij},\, i,j=e,\mu,\tau$
respect the Cheng-Sher scenerio \cite{Sher} and, therefore, the
couplings with the indices $i,j=e,\mu $ are small compared to the
ones with the indices $i=\tau\,, j=e,\mu,\tau$, since the strength
of these couplings are related to the masses of leptons denoted by
the indices of them. This forces us to study the $\tau \mu$ output
in the above processes.
\item For the coupling $\bar{\xi}^{E}_{N,\tau \mu}$ the numerical
values ($(1-10)\, GeV$) are taken by respecting the predicted
upper limit $30\, GeV$ (see \cite{Iltananomuon}) which is obtained
by using the experimental uncertainty, $10^{-9}$, in the
measurement of the muon anomalous magnetic moment
\cite{muonanmagmomexp}.
\item For the coupling $\bar{\xi}^{U}_{N,tc}$ we use the
constraint region obtained by restricting the Wilson coefficient
$C_7^{eff}$, which is the effective coefficient of the operator
$O_7 = \frac{e}{16 \pi^2}\, \bar{s}_{\alpha}\, \sigma_{\mu \nu}
\,(m_b R + m_s L)\, b_{\alpha}\, {\cal{F}}^{\mu \nu}$ (see
\cite{Alil1} and references therein), in the range $0.257 \leq
|C_7^{eff}| \leq 0.439$. Here upper and lower limits were
calculated using the CLEO measurement \cite{cleo2}
\begin{eqnarray}
BR (B\rightarrow X_s\gamma)= (3.15\pm 0.35\pm 0.32)\, 10^{-4} \,\,
, \label{br2}
\end{eqnarray}
and all possible uncertainties in the calculation of $C_7^{eff}$
\cite{Alil1}. The above restriction ensures to get upper and lower
limits for  $\bar{\xi}^{U}_{N,tt}$ and also for
$\bar{\xi}^{U}_{N,tc}$ (see \cite{Alil1} for details). In our
numerical calculations, we choose the upper limit for
$C_7^{eff}>0$, fix $\bar{\xi}^{D}_{N,bb}=30\,m_b$ and take
$\bar{\xi}^{U}_{N,tc}\sim 0.01\, \bar{\xi}^{U}_{N,tt}\sim 0.45\,
GeV$, respecting the constraints mentioned.
\end{itemize}
For the Higgs masses $m_{h^0}$ and $m_{A^0}$, we used the
numerical values $m_{h^0} = 85\, GeV$ and $m_{A^0} = 90\, GeV$. We
respect the appropriate region obtained by using the direct Higgs
boson searches and indirect limits coming from the SM
measurements, namely, $m_{h^0} > 55\, GeV$ and $m_{A^0} > 63\,
GeV$ where production of $h^0\, A^0$ is kinematically allowed at
LEP2 which has center of mass energy $~ 200 \, GeV$ (see
\cite{OpalColl}).

The addition of extra dimensions that are felt by the new Higgs
doublet results in new contributions to the decay widths of the
processes. In the case of the UEDs where all fields experience the
extra dimensions, the tree level particle-particle-KK mode
interactions are forbidden since the KK number  at each vertex
should be conserved. This leads to the non-zero contributions due
to the extra dimensions at least at one loop level and they are
suppressed. However, in the case of NUEDs, there is no need for
the conservation of KK modes at each vertex and the tree level
fermion-fermion-scalar field KK mode interaction is permitted. In
our case, the fields $h^0 \, (A^0)$ feel the extra dimensions and
their KK modes are responsible for the additional contributions
after the compactification. Our calculations are based on such
vertices and the assumption that the new Yukawa couplings existing
for the KK modes of $h^0 \, (A^0)$ are the same as the ones
existing in the zero-mode case. There is one more parameter $R$,
which is the size of the extra dimension, emerging after the
compactification and its restriction has been studied in various
works (see \cite{Antoniadis2} for example). Notice that we use a
broad range for the compactification scale $1/R$, $\,100\,GeV \leq
1/R \leq 5000\, GeV$ and present the $\Gamma$ of Higgs decays for
$1/R \leq 1000\, GeV$ since they are weakly sensitive the scale
$1/R$ for $1/R > 1000\, GeV$.

In our work, we investigate  the LFV  $t\rightarrow c \, l_i^-
l_j^+ $ and  $H^0 \rightarrow h^0 (A^0) l_i^- l_j^+$ decays in the
model III, where the new Higgs doublet and the SM gauge fields
feel extra dimension and we take  $\tau$, $\mu$ for the lepton
flavors $l_i,\, l_j$ since the Yukawa couplings, and, therefore,
the decay widths, for other pairs are highly suppressed. Here we
choose a single spatial extra dimension and, then, two spatial
extra dimensions. In the case of two spatial extra dimensions the
compactification is done  on orbifold $(S^1\times S^1)/Z_2$ and we
assume that each circle has the same radius $R$. In contrast to a
single extra dimension, the convergence of the KK sum should be
examined for two extra dimensions. In our numerical calculations,
we get convergent series for the considered range of the
compactification scale and make a rough estimate for this sum. The
numerical calculations show that the quark decay $t\rightarrow c
\, l_i^- l_j^+ $ is not sensitive to the extra dimensions,
however, the Higgs decays $H^0 \rightarrow h^0 (A^0) l_i^- l_j^+$
are sensitive, especially, to two extra dimensions.

In Fig. \ref{gammatcl1l2RExdim12}, we present the compactification
scale $1/R$ dependence of the ratio $r=\frac{\Gamma_1 }{\Gamma_0}
(\frac{\Gamma_2 }{\Gamma_0})$ for the  $t\rightarrow c (\tau^+
\mu^- + \tau^- \mu^+)$ decay, for $m_{h^0}=85\, GeV$,
$m_{A^0}=90\, GeV$, $\bar{\xi}^{D}_{N,\tau\mu}=10\, GeV$. Here
$\Gamma_0\, (\Gamma_1, \Gamma_2)$ is the decay width of the
process under consideration without extra dimension (a single
extra dimension contribution to the decay width, two extra
dimensions contribution to the decay width). The solid (dashed)
line represents the ratio $r=\frac{\Gamma_1}{\Gamma_0}
\,(\frac{\Gamma_2 }{\Gamma_0})$. This figure shows that the
contribution of the extra dimensions is suppressed for the large
values of the scale $1/R \geq 200 \, GeV$, especially for a single
extra dimension case. For two extra dimensions,  there is almost
two order enhancement in the ratio compared to the one obtained
including only one extra dimension. However, this effect is
$0.1\,\%$ of the one which is obtained without inclusion of extra
dimension and, therefore, the extra dimension contribution in this
FV decay is negligible, at least, up to two extra dimensions.

Fig. \ref{RatioH0h0A0l1l2ExtrOrdin} is devoted to the
compactification scale $1/R$ dependence of the ratio
$r=\frac{\Gamma_1 }{\Gamma_0} (\frac{\Gamma_2 }{\Gamma_0})$ for
the LFV Higgs  $H^0 \rightarrow h^0  (\tau^+ \mu^- + \tau^-
\mu^+)$ and $H^0 \rightarrow A^0  (\tau^+ \mu^- + \tau^- \mu^+)$
decays, for $m_{h^0}=85\, GeV$, $m_{A^0}=90\, GeV$,
$\bar{\xi}^{D}_{N,\tau\mu}=10\, GeV$. Here $\Gamma_0\, (\Gamma_1,
\Gamma_2)$ is the decay width of these processes without extra
dimension (a single extra dimension contribution to the decay
width, two extra dimensions contribution to the decay width). The
solid (dashed) line represents the ratio $r=\frac{\Gamma_1
}{\Gamma_0}$ for $H^0 \rightarrow h^0 (\tau^+ \mu^- + \tau^-
\mu^+)$ ($H^0 \rightarrow A^0  (\tau^+ \mu^- + \tau^- \mu^+)$)
decay and the small dashed (dotted) line represents the ratio
$r=\frac{\Gamma_2 }{\Gamma_0}$ for $H^0 \rightarrow h^0 (\tau^+
\mu^- + \tau^- \mu^+)$ ($H^0 \rightarrow A^0  (\tau^+ \mu^- +
\tau^- \mu^+)$) decay. This figure shows that the contribution of
the extra dimension is at the order of $1 \%$ of the one without
extra dimension for the large values of the scale $1/R \geq 200 \,
GeV$ and slightly larger for the $A^0$ output. In the case of two
extra dimensions, the ratio is almost one and the contribution due
to the two extra dimensions is comparable with the one without
extra dimension. This is an interesting result since these Higgs
decays are sensitive to higher dimensions and, with the more
accurate measurements, it would be possible to check the effects
of extra dimensions and to get a valuable information  about the
compactification scale.

Now, we would like to examine the effects of extra dimensions on
these Higgs decays in detail.

Fig. \ref {gammah0ksiRExdim1} (\ref {gammaA0ksiRExdim1})
represents the compactification scale $1/R$ dependence of the
decay width $\Gamma$ of $H^0 \rightarrow h^0  (\tau^+ \mu^- +
\tau^- \mu^+)$ ($H^0 \rightarrow A^0 (\tau^+ \mu^- + \tau^-
\mu^+)$) decay, for $m_{H^0}=150\, GeV$, $m_{h^0}=85\, GeV$,
$m_{A^0}=90\, GeV$, and three different values of the coupling
$\bar{\xi}^{D}_{N,\tau\mu}$. The solid (dashed, small dashed)
line:curve represents the $\Gamma$ for
$\bar{\xi}^{D}_{N,\tau\mu}=1\, GeV$ ($5\, GeV, 10\, GeV$)
without:with the inclusion of a single extra dimension. It is
shown that the $\Gamma$ of $H^0 \rightarrow h^0  (\tau^+ \mu^- +
\tau^- \mu^+)$ ($H^0 \rightarrow A^0 (\tau^+ \mu^- + \tau^-
\mu^+)$) decay is of the order of the magnitude of $10^{-5}\, GeV$
($10^{-5}\, GeV$) for the coupling $\bar{\xi}^{D}_{N,\tau\mu}=10\,
GeV$ and it enhances almost $20\, \%$ ($30\, \%$) with the
inclusion of a single extra dimension, in the range of the
compactification scale $200\, GeV\geq 1/R \geq 300\, GeV$.

In Fig. \ref {gammah0mH0RExdim1} (\ref {gammaA0mH0RExdim1}) we
present the compactification scale $1/R$ dependence of the decay
width $\Gamma$ of $H^0 \rightarrow h^0  (\tau^+ \mu^- + \tau^-
\mu^+)$ ($H^0 \rightarrow A^0 (\tau^+ \mu^- + \tau^- \mu^+)$)
decay, for $m_{h^0}=85\, GeV$, $m_{A^0}=90\, GeV$,
$\bar{\xi}^{D}_{N,\tau\mu}=10\, GeV$ and three different values of
the mass $m_{H^0}$. The solid (dashed, small dashed) line:curve
represents the $\Gamma$ for $m_{H^0}=100\, GeV$ ($150\, GeV, 170\,
GeV$) without:with the inclusion of a single extra dimension. It
is shown that for the large values of the Higgs mass  $m_{H^0}$
the $\Gamma$ of $H^0 \rightarrow h^0 (\tau^+ \mu^- + \tau^-
\mu^+)$ ($H^0 \rightarrow A^0 (\tau^+ \mu^- + \tau^- \mu^+)$)
decay is of the order of the magnitude of $10^{-2}\, GeV$
($10^{-4}\, GeV$) and the sensitivity of the extra dimension
becomes smaller with the increasing values of the Higgs masses.

Fig. \ref {gammah0mh00RExdim1} (\ref {gammaA0mA0RExdim1}) is
devoted to the compactification scale $1/R$ dependence of the
decay width $\Gamma$ of $H^0 \rightarrow h^0  (\tau^+ \mu^- +
\tau^- \mu^+)$ ($H^0 \rightarrow A^0 (\tau^+ \mu^- + \tau^-
\mu^+)$) decay, for $m_{H^0}=150\, GeV$, $m_{A^0}=90\, GeV$
($m_{h^0}=80\, GeV$) $\bar{\xi}^{D}_{N,\tau\mu}=10\, GeV$ and
three different values of the mass $m_{h^0}$ ($m_{A^0}$). The
solid (dashed, small dashed) line:curve represents the $\Gamma$
for $m_{h^0}=75\, GeV$ ($80\, GeV, 85\, GeV$)  ($m_{A^0}=90\, GeV$
($100\, GeV, 120\, GeV$)) without:with the inclusion of a single
extra dimension. Here we see that the increase in the mass values
$m_{h^0}$ ($m_{A^0}$) causes  the decay width to decrease and the
sensitivity to the single extra dimension to increase .

Now we would like to present the results briefly:

\begin{itemize}
\item  For the $t\rightarrow c (\tau^+ \mu^- + \tau^- \mu^+)$
decay,  the contribution of the extra dimensions is small for the
large values of the scale $1/R \geq 200 \, GeV$, especially for a
single extra dimension case. In the case of two extra dimensions
the additional contribution is almost two order larger  compared
to the one obtained for a single extra dimension. In any case,
this effect is $0.1\,\%$ of the contribution which is obtained
without inclusion of extra dimension and, therefore, the extra
dimension contribution is negligible in this FV decay, at least,
up to two extra dimensions.
\item The decay widths of LFV $H^0 \rightarrow h^0  (\tau^+ \mu^-
+ \tau^- \mu^+)$ and $H^0 \rightarrow A^0  (\tau^+ \mu^- + \tau^-
\mu^+)$ decays are sensitive to the extra dimensions. The new
effect coming from a single extra dimension is of the order of $1
\%$ of the contribution obtained without extra dimension for the
large values of the scale $1/R \geq 200 \, GeV$. In the case of
two extra dimensions the new effects are almost comparable with
the one  obtained without extra dimension.
\end{itemize}

As a final comment, the Higgs decays $H^0 \rightarrow h^0  (\tau^+
\mu^- + \tau^- \mu^+)$ and $H^0 \rightarrow A^0  (\tau^+ \mu^- +
\tau^- \mu^+)$ are sensitive to the extra dimensions and with the
more accurate future measurements it would be possible to check
effects of extra dimensions and predict valuable information about
the compactification scale.
\section{Acknowledgement}
This work has been supported by the Turkish Academy of Sciences in
the framework of the Young Scientist Award Program.
(EOI-TUBA-GEBIP/2001-1-8)
\newpage
\begin{figure}[htb]
\vskip 2.0truein \centering \epsfxsize=2.8in
\leavevmode\epsffile{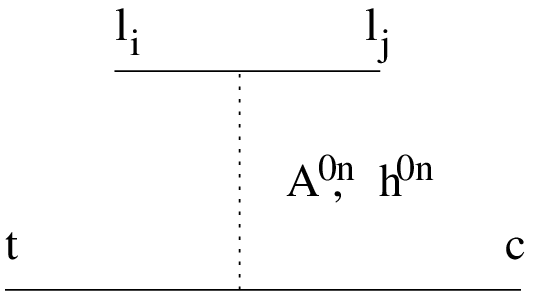} \vskip 1.0truein \caption[]{Tree
level diagrams contributing to the decay $t\rightarrow c \, l_i^-
l_j^+$. Dotted lines represent the $h^{0 n},A^{0 n}$ fields where
$n=0,1,2,...$} \label{fig1}
\end{figure}
\begin{figure}[htb]
\vskip 2.0truein \centering \epsfxsize=6.8in
\leavevmode\epsffile{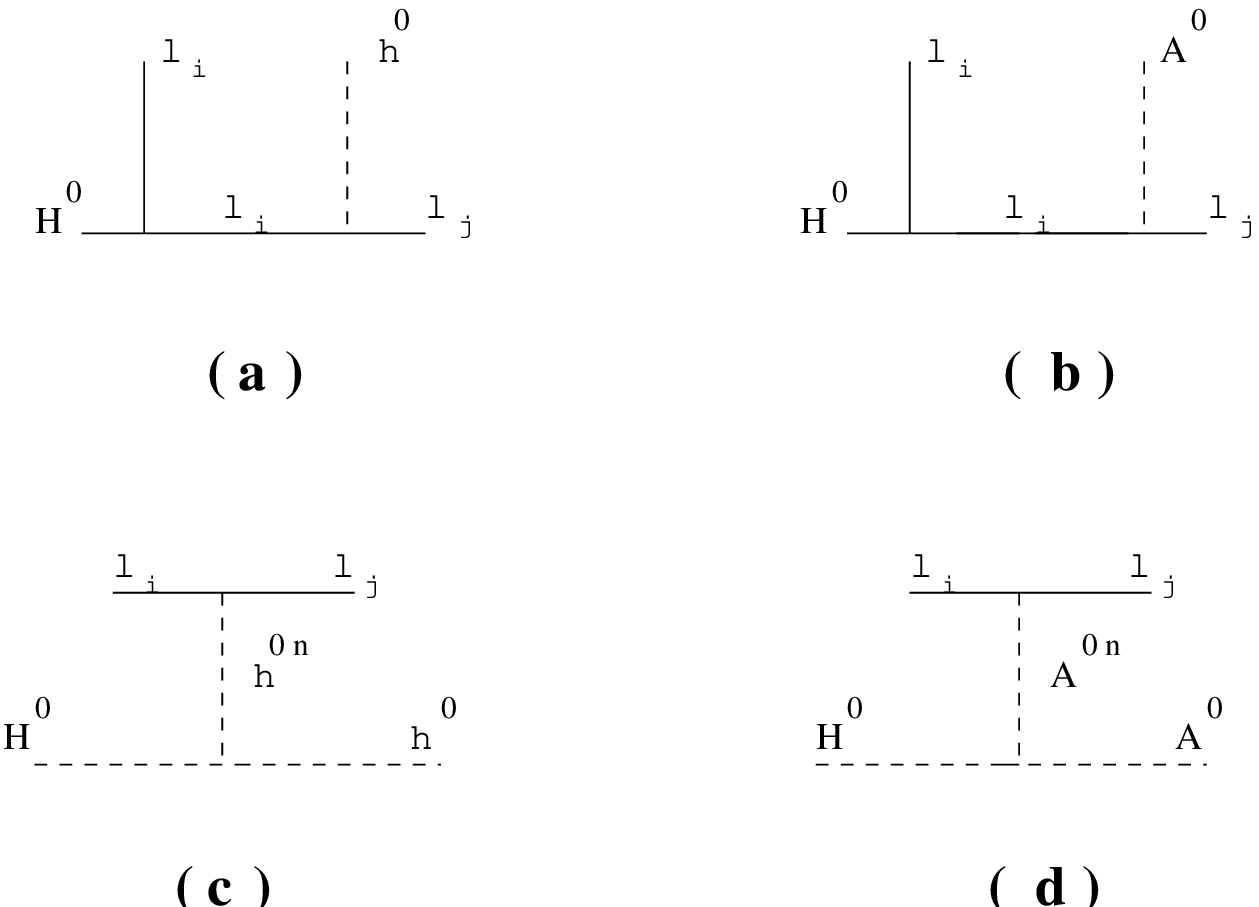} \vskip 1.0truein \caption[]{Tree
level diagrams contributing to $\Gamma (H^0\rightarrow h^0 (A^0)\,
l_i^- l_j^+)$, $i=e,\mu,\tau$ decay  in the model III version of
2HDM. Solid lines represent leptons, dashed lines represent the
$H^0$, $h^0$ and $A^0$ fields, where $n=0,1,2,...$} \label{fig2}
\end{figure}
\newpage
\begin{figure}[htb]
\vskip -3.0truein \centering \epsfxsize=6.8in
\leavevmode\epsffile{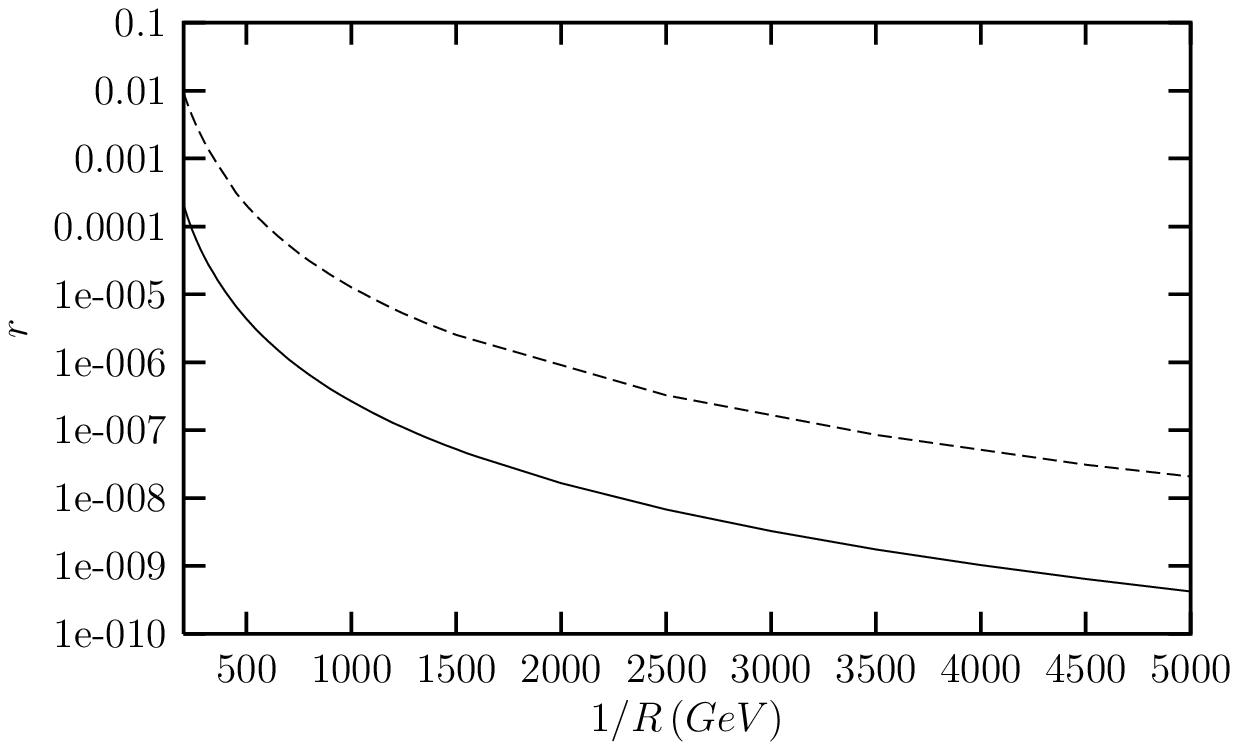} \vskip -3.0truein
\caption[]{The compactification scale $1/R$ dependence of the
ratio $r=\frac{\Gamma_1 }{\Gamma_0} (\frac{\Gamma_2 }{\Gamma_0})$
for the decay $t\rightarrow c (\tau^+ \mu^- + \tau^- \mu^+)$ for
$m_{h^0}=85\, GeV$, $m_{A^0}=90\, GeV$,
$\bar{\xi}^{D}_{N,\tau\mu}=10\, GeV$. The solid (dashed) line
represents the ratio $r=\frac{\Gamma_1}{\Gamma_0}
\,(\frac{\Gamma_2 }{\Gamma_0})$.} \label{gammatcl1l2RExdim12}
\end{figure}
\begin{figure}[htb]
\vskip -3.0truein \centering \epsfxsize=6.8in
\leavevmode\epsffile{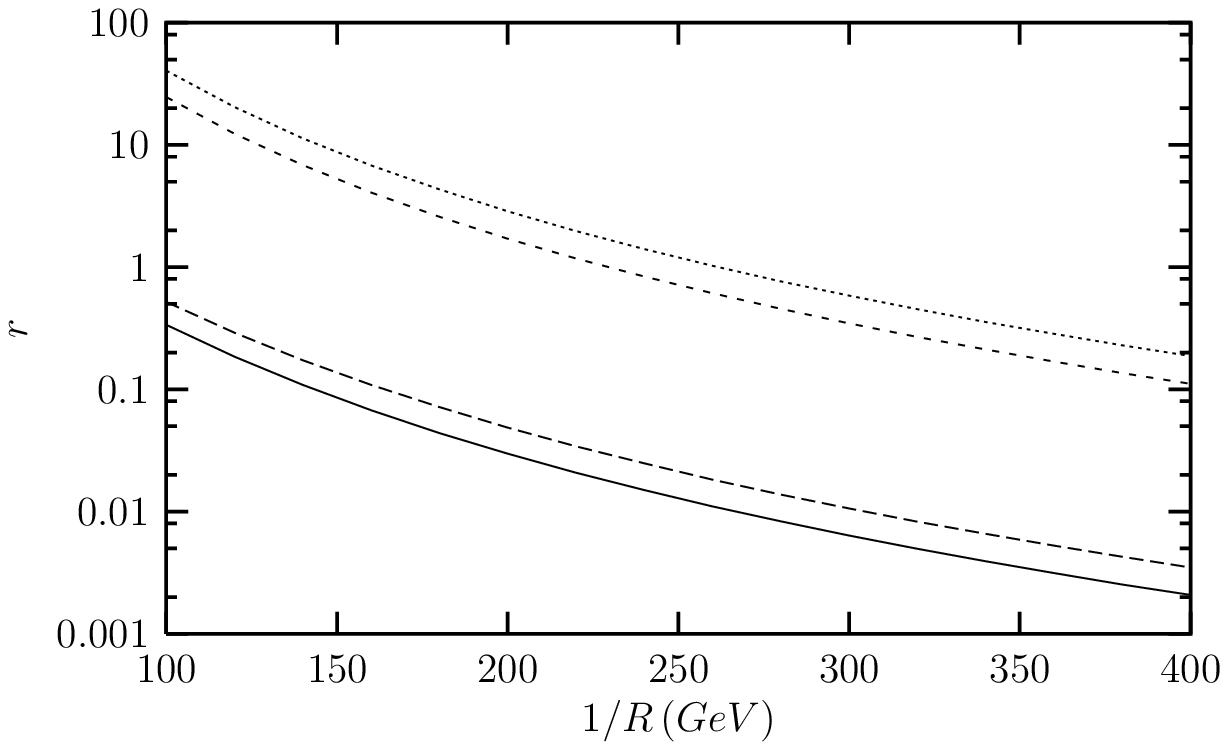} \vskip
-3.0truein \caption[]{The compactification scale $1/R$ dependence
of the ratio $r=\frac{\Gamma_1 }{\Gamma_0} (\frac{\Gamma_2
}{\Gamma_0})$ for the LFV Higgs  $H^0 \rightarrow h^0  (\tau^+
\mu^- + \tau^- \mu^+)$ and $H^0 \rightarrow A^0  (\tau^+ \mu^- +
\tau^- \mu^+)$ decays for $m_{h^0}=85\, GeV$, $m_{A^0}=90\, GeV$,
$\bar{\xi}^{D}_{N,\tau\mu}=10\, GeV$. The solid (dashed) line
represents the ratio $r=\frac{\Gamma_1 }{\Gamma_0}$ for $H^0
\rightarrow h^0 (A^0) (\tau^+ \mu^- + \tau^- \mu^+)$ decay and the
small dashed (dotted) line represents the ratio $r=\frac{\Gamma_2
}{\Gamma_0}$ for $H^0 \rightarrow h^0 (A^0) (\tau^+ \mu^- + \tau^-
\mu^+)$ decay.} \label{RatioH0h0A0l1l2ExtrOrdin}
\end{figure}
\begin{figure}[htb]
\vskip -3.0truein \centering \epsfxsize=6.8in
\leavevmode\epsffile{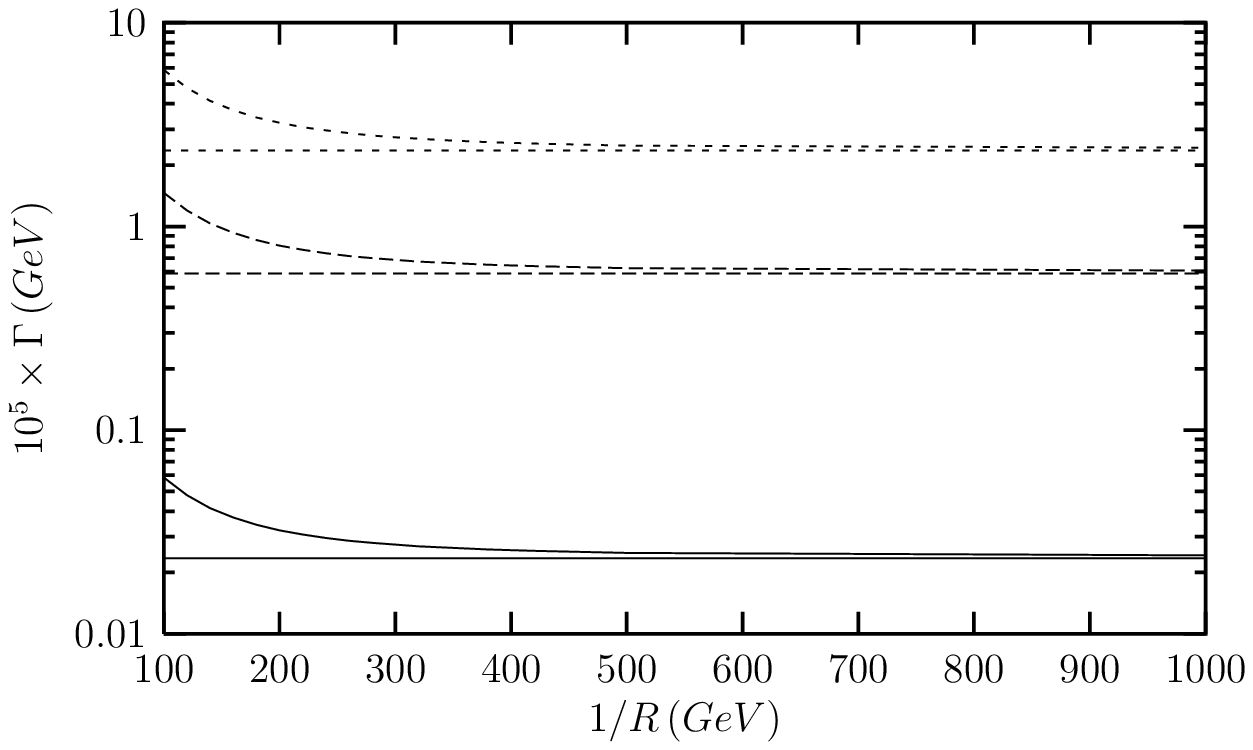} \vskip -3.0truein
\caption[]{The compactification scale $1/R$ dependence of the
decay width $\Gamma$ of $H^0 \rightarrow h^0  (\tau^+ \mu^- +
\tau^- \mu^+)$  decay, for $m_{H^0}=150\, GeV$, $m_{h^0}=85\,
GeV$, $m_{A^0}=90\, GeV$, and three different values of the
coupling $\bar{\xi}^{D}_{N,\tau\mu}$. The solid (dashed, small
dashed) line:curve represents the $\Gamma$ for
$\bar{\xi}^{D}_{N,\tau\mu}=1\, GeV$ ($5\, GeV, 10\, GeV$)
without:with the inclusion of a single extra dimension.}.
\label{gammah0ksiRExdim1}
\end{figure}
\begin{figure}[htb]
\vskip -3.0truein \centering \epsfxsize=6.8in
\leavevmode\epsffile{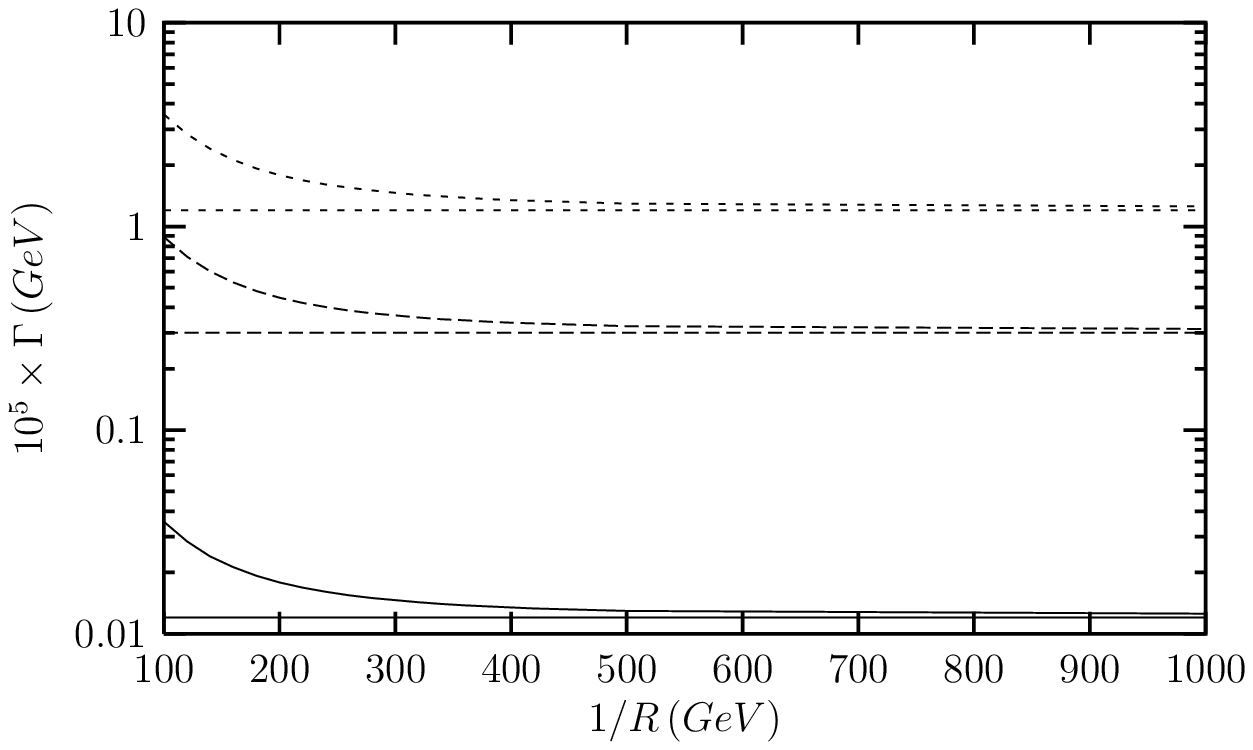} \vskip -3.0truein
\caption[]{The same as Fig. \ref{gammah0ksiRExdim1} but for $H^0
\rightarrow A^0  (\tau^+ \mu^- + \tau^- \mu^+)$  decay.}
\label{gammaA0ksiRExdim1}
\end{figure}
\begin{figure}[htb]
\vskip -3.0truein \centering \epsfxsize=6.8in
\leavevmode\epsffile{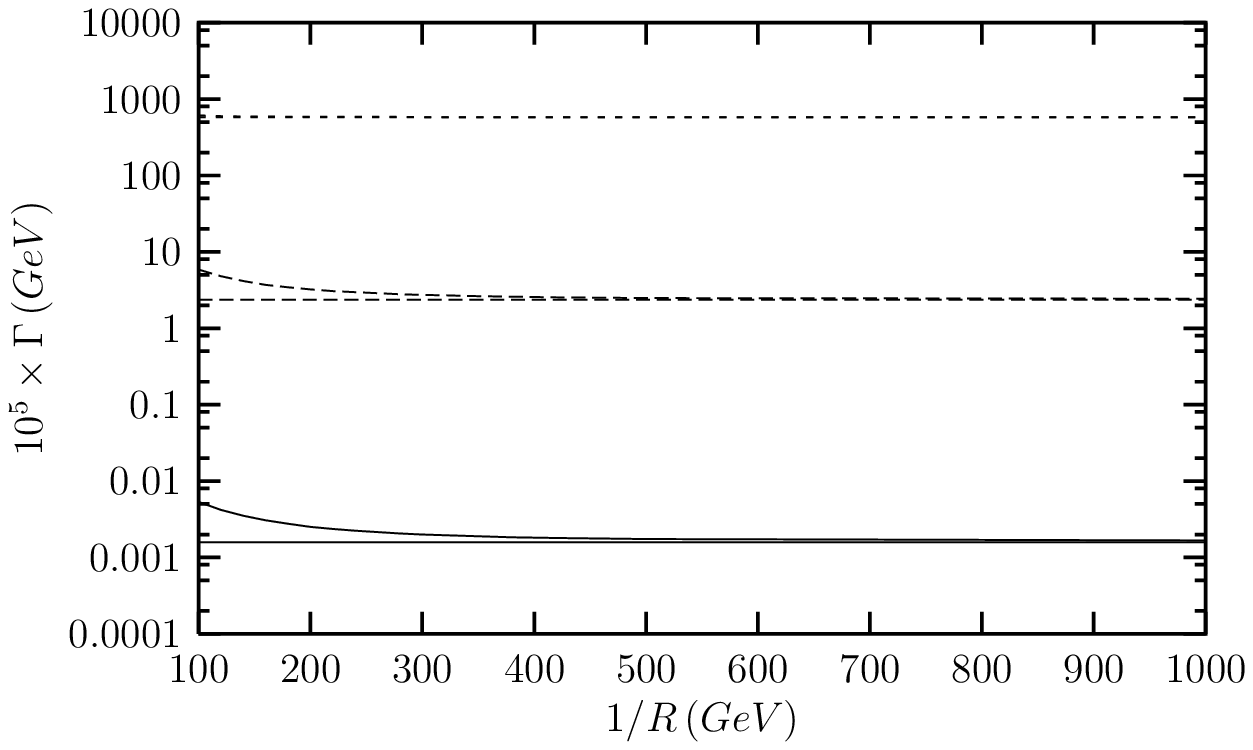} \vskip -3.0truein
\caption[]{The compactification scale $1/R$ dependence of the
decay width $\Gamma$ of $H^0 \rightarrow h^0  (\tau^+ \mu^- +
\tau^- \mu^+)$ decay, for $m_{h^0}=85\, GeV$, $m_{A^0}=90\, GeV$,
$\bar{\xi}^{D}_{N,\tau\mu}=10\, GeV$ and three different values of
the mass $m_{H^0}$. The solid (dashed, small dashed) line:curve
represents the $\Gamma$ for $m_{H^0}=100\, GeV$ ($150\, GeV, 170\,
GeV$) without:with the inclusion of a single extra dimension.}
\label{gammah0mH0RExdim1}
\end{figure}
\begin{figure}[htb]
\vskip -3.0truein \centering \epsfxsize=6.8in
\leavevmode\epsffile{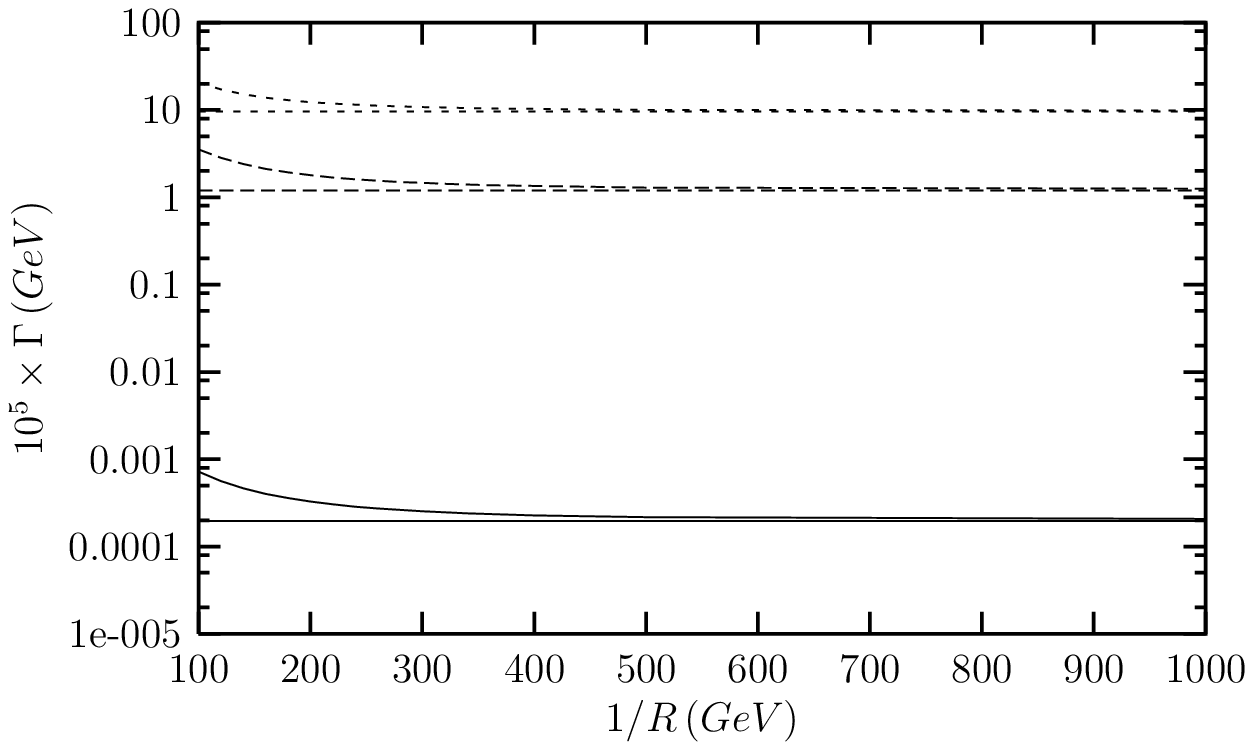} \vskip -3.0truein
\caption[]{The same as Fig. \ref{gammah0mH0RExdim1} but for $H^0
\rightarrow A^0  (\tau^+ \mu^- + \tau^- \mu^+)$  decay.}
\label{gammaA0mH0RExdim1}
\end{figure}
\begin{figure}[htb]
\vskip -3.0truein \centering \epsfxsize=6.8in
\leavevmode\epsffile{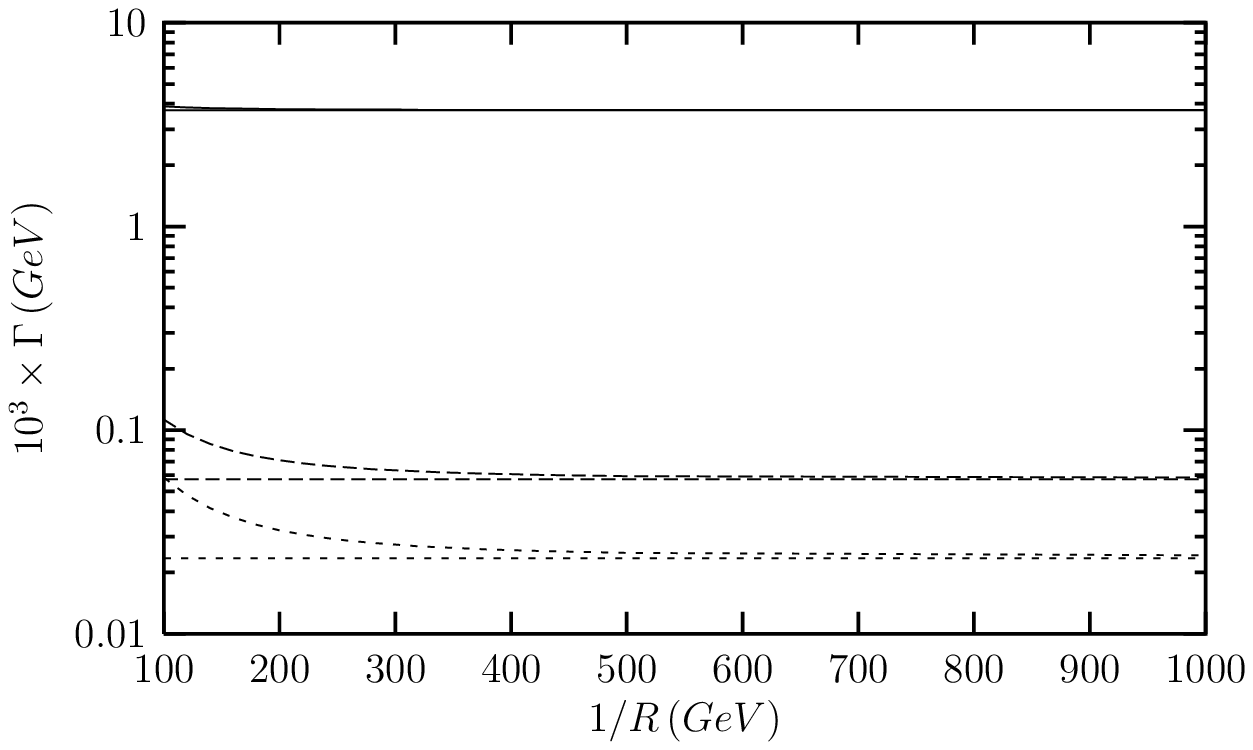} \vskip -3.0truein
\caption[]{The compactification scale $1/R$ dependence of the
decay width $\Gamma$ of $H^0 \rightarrow h^0  (\tau^+ \mu^- +
\tau^- \mu^+)$ decay, for $m_{H^0}=150\, GeV$, $m_{A^0}=90\, GeV$
 $\bar{\xi}^{D}_{N,\tau\mu}=10\, GeV$ and
three different values of the mass $m_{h^0}$. The solid (dashed,
small dashed) line:curve represents the $\Gamma$ for $m_{h^0}=75\,
GeV$ ($80\, GeV, 85\, GeV$)   without:with the inclusion of a
single extra dimension. } \label{gammah0mh00RExdim1}
\end{figure}
\begin{figure}[htb]
\vskip -3.0truein \centering \epsfxsize=6.8in
\leavevmode\epsffile{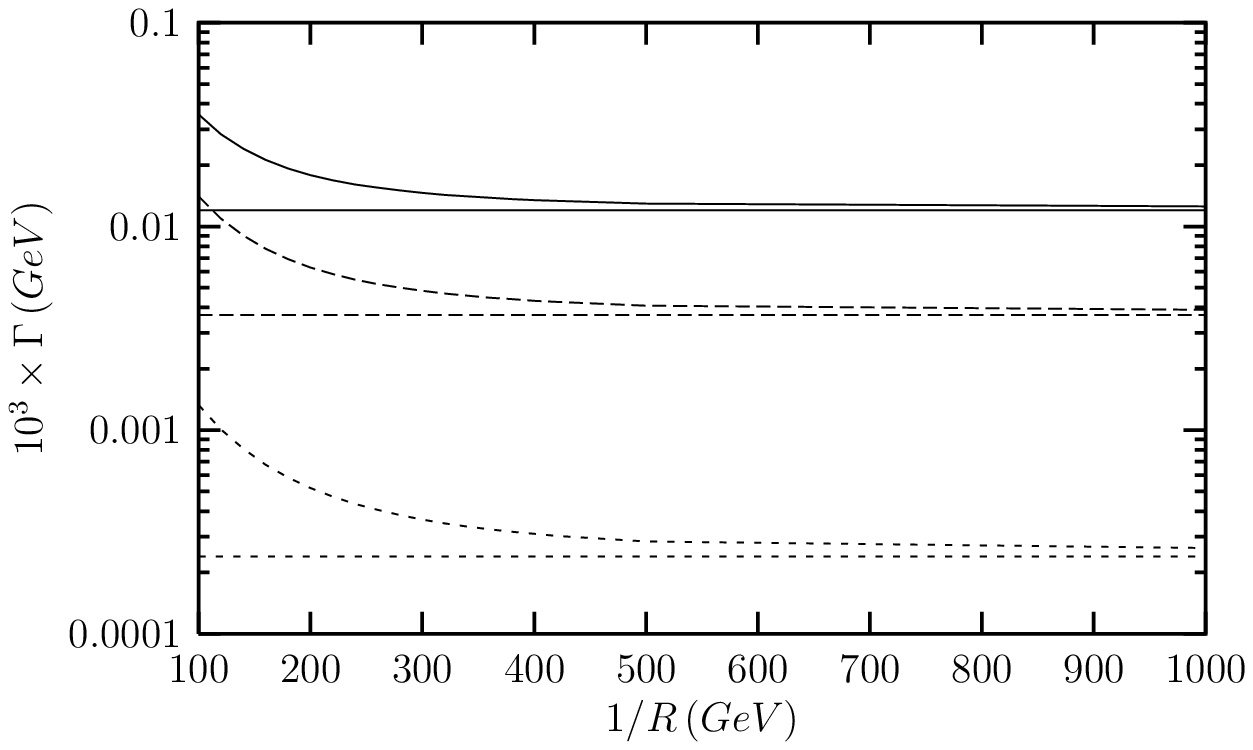} \vskip -3.0truein
\caption[]{The compactification scale $1/R$ dependence of the
decay width $H^0 \rightarrow A^0 (\tau^+ \mu^- + \tau^- \mu^+)$
decay, for $m_{H^0}=150\, GeV$, $m_{h^0}=80\, GeV$,
$\bar{\xi}^{D}_{N,\tau\mu}=10\, GeV$ and three different values of
the mass $m_{A^0}$. The solid (dashed, small dashed) line:curve
represents the $\Gamma$ for   $m_{A^0}=90\, GeV$ ($100\, GeV,
120\, GeV$) without:with the inclusion of a single extra
dimension.} \label{gammaA0mA0RExdim1}
\end{figure}
\end{document}